\documentclass[twocolumn,prx]{revtex4-2}
\usepackage[utf8]{inputenc}
\usepackage{amsmath}
\usepackage{amssymb}
\usepackage{bm}
\usepackage{graphicx}
\usepackage{color}

\begin{document}

\title{Enhancement of the Kondo effect in a quantum dot formed in a full-shell nanowire.}
\author{Aleksandr E. Svetogorov}
\author{Daniel Loss}
\author{Jelena Klinovaja}
\affiliation{Department of Physics, University of Basel, Klingelbergstrasse 82, CH-4056 Basel, Switzerland}

\date{\today}
\begin{abstract}
We analyze results of a recent experiment [D. Razmadze et al., \textit{Phys. Rev. Lett.}, \textbf{125}, 116803 (2020)] on transport through a quantum dot between two full-shell nanowires and show that the observed effects are caused by the Kondo effect enhancement due to a nontrivial geometry (magnetic flux in a full-shell nanowire) rather than the presence of Majorana bound states. Moreover, we propose that such a setup presents a unique and convenient system to study the competition between superconductivity and the Kondo effect and has significant advantages in comparison to other known approaches, as the important parameter is controlled by the magnetic flux through the full-shell nanowire, which can be significantly varied with small changes of magnetic field, and does not require additional gates. This competition is of fundamental interest as it results in a quantum phase transition between an unscreened doublet and a many-body Kondo singlet ground states of the system.
\end{abstract}
\maketitle

\section{Introduction}
Semiconducting nanowires with full superconducting shell were recently introduced as possible realizations of topological superconductors, which may host Majorana bound states (MBSs)~\cite{Glazman_Marcus2020}. MBSs in turn have non-Abelian statistics, which could be exploited to develop a topologically protected qubit~\cite{Fisher2011,Alicea2016}. In nanowires with a thin shell Little-Parks effect~\cite{Parks1962} results in modulation of the superconducting order parameter with the applied magnetic flux. In case of a thin nanowire (with a diameter smaller than the superconducting coherence length)  the Little parks effect is destructive resulting in a lobe structure of the order parameter as a function of the flux~\cite{Zharkov1980,Marcus2020_1}. The idea of combining effectively one-dimensional superconductors with a vortex is already fascinating by itself: the vortex can induce a  phase winding of the superconducting order parameter and result in nontrivial properties, such as the well-known Caroli–de Gennes–Matricon bound states in 2D case~\cite{Caroli1964}. Further experiments with full-shell nanowires were performed~\cite{Sabonis2020,Nygard2021,Marcus2021}, including those where the observed zero-bias anomalies were shown to have non-topological nature~\cite{Katsaros2021}. However, a recent experiment~\cite{Marcus2020} with a gate-controlled quantum dot (QD) in a full-shell hybrid interferometer showed non-trivial features in addition to the zero-bias peak. Without a magnetic flux (zeroth lobe) through the shell a change of occupation of the dot (from even to odd) leads to a change of a sign of the supercurrent through the dot, which is known as $0-\pi$ phase transition~\cite{Matveev1989,Kouwenhoven2006,Bouchiat2006,Novotny2007}. However, with a flux around one magnetic flux quantum threading the superconducting shell (first lobe) the effective Josephson junction seems to stay in the $0$ phase even for the odd occupation of the QD, 
in agreement
 with theoretical predictions~\cite{Zheng2015,Oppen2017,Fu2018,Black-Schaffer2019,Flensberg2020} for a Josephson junction based on QD between two topological superconductors hosting MBSs. Such peculiar behaviour could be seen as  
 evidence for the presence of MBSs in the system. Nevertheless, the authors were not completely satisfied with the interpretation, as they could not explain some of the observed features, such as an enhancement of the supercurrent in the odd state in the first lobe and no signs of fractional Josephson effect expected in the presence of MBSs were observed. 

In this paper we argue that the results of this experiment can be interpreted as an enhancement of the Kondo effect~\cite{Glazman2001,Schonenberger2002,Belzig2004,Florens2009,Aguado2015} in the first lobe, including an   enhancement of the supercurrent. It was predicted theoretically~\cite{Yeyati2003,Belzig2004,Barash2005,Meden2008} and shown experimentally~\cite{Schonenberger2009,Deblock2015,Franceschi2012} that if the Kondo effect can develop on an odd-occupied QD between superconducting leads, the ground state is a many-body Kondo singlet instead of a doublet (unscreened electron), which restores the $0$ phase behaviour and enhances the critical supercurrent. The phase transition is determined by the ratio of two energy scales: the Kondo temperature $T_K$ and superconducting gap $\Delta$. In this paper we show that if the superconducting order parameter acquires a phase winding around the shell, the relevant energy scale is not the absolute value of the gap $|\Delta|$ but an effective gap which is significantly suppressed due to destructive interference, which in turn leads to
the enhancement of the Kondo effect. Our interpretation is further supported by the fact that a bright zero-bias feature and no $\pi$ phase is observed at the closing of the zeroth lobe (around half the flux quantum), where no topological superconductivity is expected.  
As a result, we can claim that due to the detailed 
experimental data provided by Razmadze et al.~\cite{Marcus2020} it is possible to establish a non-trivial effect of the phase winding on the coherent transport and the ground state properties of a QD between two full-shell nanowires, which has not been predicted before. Further experimental studies of the effect can enrich the understanding of the underlying fundamental physics as only elaborate numerical approaches have been developed to quantitatively capture the QD-based Josephson junction behaviour in the competition regime~\cite{Satori1993,Ohashi2000,Belzig2004,Hewson2007,Meden2008} (for review see~\cite{Meden2019}). 

\section{Model}
\begin{figure}
\includegraphics[width=8cm]{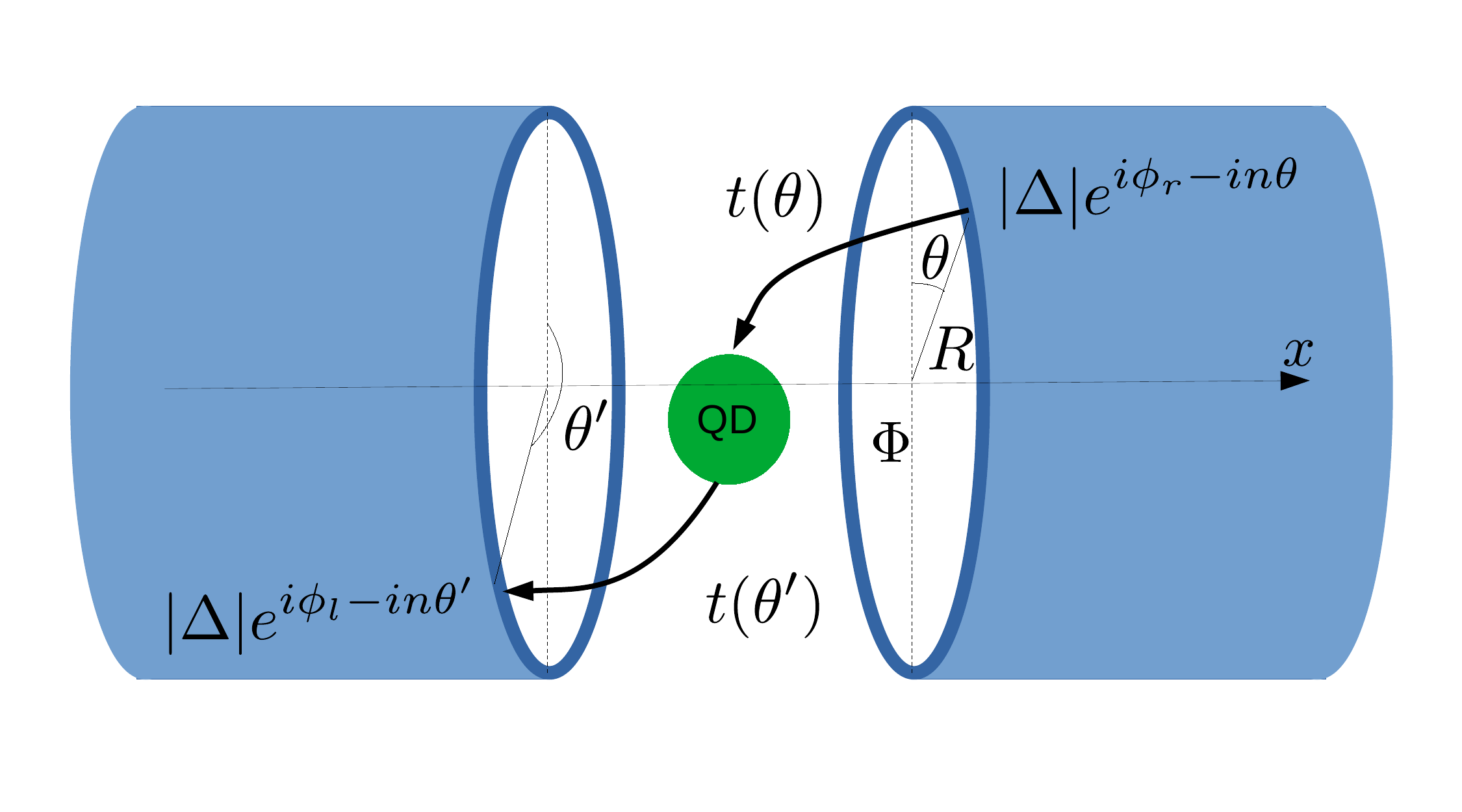}
\caption{\label{fig:SQDS}
Schematic representation of Cooper-pair tunneling through a QD electron state (green sphere) between two hollow-cylinder superconductors (blue) representing the shells and accumulation layers of underlying nanowires. The tunneling amplitude $t(\theta)$ (between the dot and each shell) is angle-dependent, as the electron wave function on the dot is shifted from the shell axis due to gating; the superconducting order parameter on the shell has angle dependence - nonzero phase winding ($n\neq0$ for large enough magnetic flux $\Phi>\Phi_0/2$).}
\end{figure}
In this paper we focus on a QD formed in an uncovered (etched superconductor) and gated region of a full-shell nanowire. Assuming that electrons in the nanowire form an accumulation layer at the boundary with the superconducting shell~\cite{Glazman_Marcus2020}, we use a hollow-shell approximation - transport through the system is mostly determined by Cooper pairs propagation along the shell and tunneling between the shells through the QD. We assume the shell to be much thinner than the magnetic penetration length and the diameter to be smaller than the superconducting coherence length, which results in no quantization of the magnetic flux through the shell but reentrant lobe structure due to the destructive Little-Parks effect: superconductivity is fully suppressed when the flux around odd half-integer multiples of a flux quantum is applied~\cite{Marcus2020, Marcus2020_1}. As a further simplification we model the shell as a hollow cylinder (in experiments it has rather a hexagonal cross-section)~\cite{Glazman_Marcus2020,Prada2022} with radius $R$, then the total magnetic flux through the shell is $\Phi=\pi R^2B$. We focus on the single-level QD limit (large level spacing), the number of electrons can be considered fixed by the applied gate due to strong Coulomb interactions on the QD (strong spatial confinement), which is a common experimental situation. The simplest model to describe such a system is the superconducting impurity Anderson (SIA) model~\cite{Anderson1961} modified by the magnetic flux though the shell~\cite{Prada2022}:
\begin{align}
&H=H_D+\sum_{s=l/r}\left(H_s+V^s_{SD}\right),\\
&H_D=\sum_{\sigma}\left(\epsilon_{d}+\sigma V_{Z}\right)d^\dagger_\sigma d_\sigma+Un_\uparrow n_\downarrow,\\
&H_s=\int dx d\theta\sum_\sigma\left[\psi^\dagger_{\sigma\theta x}\left(\frac{p^2}{2m_s}-\mu_s\right)\psi_{\sigma\theta x}\right.\notag \\ 
&+\left.\Delta_s(\Phi) e^{-in\theta+i\phi_s}\psi^\dagger_{\sigma\theta x}\psi^\dagger_{-\sigma\theta x}+\mathrm{h.c.}\right],\\
&V_{SD}^s=\int d\theta\sum_\sigma t_s(\theta)\psi^\dagger_{\sigma\theta x}d_\sigma/(2\pi)+\mathrm{h.c.}
\end{align}
Here $\epsilon_d$ is the bare dot energy level, which is controlled by gate voltage, $V_Z=\mu_BgB/2$ is the Zeeman field ($g$ is Land\'e g-factor and $\mu_B$ is Bohr magneton), $\sigma$ corresponds to electron spin up/down state ($\pm1$ if it is a coefficient, $\uparrow/\downarrow$ if it is an index); $d^\dagger_\sigma$ is the QD electron creation operator (with spin $\sigma$), $n_\sigma=d^\dagger_\sigma d_\sigma$ is the occupation number operator, $U$ is the charging energy (which is the largest energy scale in the system so that the single-level approximation is valid); $\psi_{\sigma\theta x}$ are the shell operators (we omit $s=l/r$ index for the left/right shell), $x$ is the coordinate along the shells, $\theta$ is the angle around the shell. The superconducting order parameter in the shells $\Delta_{l/r}(\Phi,\theta)=\left|\Delta(\Phi)_{l/r}\right|e^{-in\theta+i\phi_{l/r}}$ depends on magnetic flux $\Phi$ as well as angle $\theta$, $\phi_{l/r}$ is the phase at the same angle $\theta=0$ to the left/right of the QD, $n$ denotes the number of phase windings defined by an integer of the ratio of magnetic flux to flux quantum ($\Phi_0=\pi/e$, we set $\hbar=1$ throughout the paper): $n=\lfloor\Phi/\Phi_0\rceil$; $p^2/(2m_s)$ and $\mu_s$ are the kinetic term and the chemical potential in the respective shells. The last term describes tunneling between the left/right shell and the QD, the tunneling amplitude is given by $t_s(\theta)/(2\pi)$ and dependends on the azimuthal angle $\theta$, as the electron state on the QD is shifted away from the shell axis due to the dot gating. We work in the zero-temperature limit (the temperature in relevant experiments is well below characteristic energies). The main difference from~\cite{Prada2022} is the presence of two superconducting shells instead of one, which adds up and introduces an additional important parameter, namely the phase difference $\phi=\phi_l-\phi_r$ between the shells (at the same angle $\theta$). The effective Hamiltonian can be obtained by integrating out the shells and introducing the dot Green function as $G(\omega)=\left[\omega-H_{eff}(\omega)\right]^{-1}$:
\begin{equation*}
H_{eff}(\omega)=H_{eff,\uparrow}(\omega)\oplus H_{eff,\downarrow}(\omega)
\end{equation*}
\begin{equation}\label{eq:hamiltonian}
H_{eff,\sigma}\left(\omega\right)=\left(\begin{array}{cc}
\epsilon_{d}+\sigma V_{z} & 0\\
0 & -\epsilon_{d}+\sigma V_{z}
\end{array}\right)+\Sigma_{\sigma}^{U}+\Sigma^{S},
\end{equation}
where the first term corresponds to the bare QD. The last two terms are self-energies from the Coulomb interaction $U$ on the dot and from the proximity effect induced by the superconducting shells, respectively. The proximity effect contributes as
\begin{equation}
\Sigma^{S}=-\frac{2\left\langle \Gamma_{S}\right\rangle }{\sqrt{\Delta^{2}-\omega^{2}}}\left(\begin{array}{cc}
\omega & \Delta_{eff}\cos\frac{\phi}{2}\\
\Delta_{eff}\cos\frac{\phi}{2} & \omega
\end{array}\right).
\end{equation}
Here $\left\langle \Gamma_{S}\right\rangle=\pi\rho_{s}\left[\intop_{-\pi}^\pi\frac{d\theta}{2\pi} t\left(\theta\right)\right]^2$ stands for the tunneling energy scale averaged over the angle $\theta$ around the shell (the contribution comes from hopping from QD to shell and back with amplitude $t(\theta)/(2\pi)$), $\rho_s$ is the density of states at the Fermi energy, $\phi$ is the phase difference between left and right shell (at fixed angle $\theta$). We assume symmetric tunneling $t_r(\theta)=t_l(\theta)=t(\theta)$ to the left/right shell, as corresponding results can be easily generalized for an asymmetric case~\cite{Novotny2017}. Even more nontrivial contribution comes in the numerator of the off-diagonal terms (see Appendix~\ref{app:effdelta}):
\begin{equation}\label{eq:deff}
\Delta_{eff}\approx(1-\delta_{0n})\left|\frac{2\Delta\intop_{-\pi}^{\pi}t\left(\theta\right)e^{i\theta n}\frac{d\theta}{2\pi}}{\intop_{-\pi}^{\pi}t\left(\theta\right)\frac{d\theta}{2\pi}}\right|+\delta_{0n}\left|\Delta\right|.
\end{equation}
One can see that for $n\neq0$ and axially symmetric tunneling ($t(\theta)=const$), $\Delta_{eff}=0$! One should note that Eq.~(\ref{eq:deff}) is approximate, it works well for weak dependence of tunneling amplitude on $\theta$; for general case see analyses in Appendix~\ref{app:effdelta}.

The effect can be interpreted as destructive interference, which can be seen from a schematic of a Cooper-pair tunneling trajectory between two superconducting hollow cylinders through a QD electronic state, Fig.~\ref{fig:SQDS}; for axially symmetric case a Cooper pair has equal probability to tunnel between all possible angle positions on superconducting shells.  If there is a phase winding around the shells, each trajectory corresponds to some phase difference $\Delta\phi\in[0,2\pi n]$, then summing over all trajectories gives exactly zero (which can be written as zero effective gap $\Delta_{eff}$). However, the effective QD state (the bound state wave-function) is rather not centred on the shell axis, as the QD is gated from one side, which results in $\theta$-dependent tunneling amplitude $t(\theta)$ and, therefore, nonzero $\Delta_{eff}$. As was discussed in~\cite{Prada2022}, the angle-dependence of tunneling is rather weak as the tunneling is determined by the tails of the wave-function under the superconducting shell, which is screening the electrical field form the gate. The most straightforward consequence of such a destructive interference is the reduction of the Josephson effect by a factor $\Delta_{eff}/\Delta$. 

However, that is not the whole story. First of all, as it was already briefly discussed, if a flux around half flux quantum is applied, superconductivity is completely suppressed due to Little-Parks effect. Second, even an S-QD-S junction without magnetic flux shows nontrivial behaviour such as $0-\pi$ phase transition. This phase transition was extensively studied theoretically, starting with the first predictions for transition induced by changing the occupation of the QD~\cite{Matveev1989,Kouwenhoven2006,Bouchiat2006,Novotny2007,Wernsdorfer2010} and followed by more sophisticated regimes, when the Kondo effect may play a significant role~\cite{Shimizu1992,Satori1993,Ohashi2000,Belzig2004,Hewson2007,Meden2008}. These effects are due to strong Coulomb interactions on the QD represented by $\Sigma_{\sigma}^{U}$ term in the effective Hamiltonian. In~\cite{Prada2022} it was calculated in Hartree-Fock-Bogoliubov approximation ~\cite{Arovas1999,Yeyati2003} (the lowest $U$-order expansion)
\begin{equation}\label{eq:ssigma}
\Sigma_{\sigma}^{U}\approx U\left(\begin{array}{cc}
\langle n_{-\sigma}\rangle & \langle d_\sigma d_{-\sigma}\rangle\\
\langle d^\dagger_\sigma d^\dagger_{-\sigma}\rangle & -\langle n_{\sigma}\rangle
\end{array}\right),
\end{equation}
which cannot capture the Kondo effect. The latter was studied with powerful numerical approaches~\cite{Satori1993,Ohashi2000,Belzig2004,Hewson2007} as of now no reliable analytical approach capable of tackling the problem in the most interesting regime of competition between the Kondo effect and superconductivity has been developed. Fully analytical methods are available only for special limiting cases such as Hartree–Fock-Bogoliubov approximation~\cite{Arovas1999,Yeyati2003} and perturbation in cotunneling~\cite{Konig2013,Paaske2015} through Yu-Shiba-Rusinov (YSR) state~\cite{Yu1965,Shiba1968,Rusinov1969} analogs for $\Delta\gg T_K$ or  slave-boson mean field approaches in the opposite limit~\cite{Zaikin2003,Vecino2003}.  Nevertheless, it is well established that if the Kondo temperature $T_K$ (characteristic energy scale) is large enough in comparison to the superconducting gap $\Delta$, the electron on a dot can form a Kondo singlet with quasiparticles in the superconductor (Kondo cloud) and, therefore, the junction stays in the $0$ phase even in the odd sector, the cotunneling process is enhanced which in turn increases the supercurrent~\cite{Hewson2007,Meden2008,Florens2009}. The Kondo temperature depends on tunneling amplitude, Coulomb interaction, and bare QD level energy~\cite{Haldane1978}:
\begin{equation}
T_K\sim\sqrt{U\Gamma}\exp\left[\frac{\pi\epsilon_d}{2\Gamma}\left(1+\frac{\epsilon_d}{U}\right) \right],\quad \Gamma=2\langle\Gamma_s\rangle.
\end{equation}
The exact proportionality factor $0.28$ is well defined only in the middle of the odd occupation region ($\epsilon_d=-U/2$)~\cite{Meden2019,Florens2020}.
An important question is how the superconducting phase winding affects this competition.

\section{Analyses of the experiment}
In this section we analyze the results of an experiment performed on a S-QD-S junction with a flux through the full-shell nanowire 
with the goal to establish topological superconductivity~\cite{Marcus2020}. Several non-trivial features were reported that could potentially indicate the presence of 
Majorana fermions. First, the differential conductance in a voltage-bias configuration was measured, then the current-phase relation (CPR) was probed in a SQUID geometry. The even-occupied regime does not show anything unexpected: the differential conductance clearly shows a gap-closing around half flux quantum due to the destructive Little-Parks effect and a gap-reopening in the first lobe (Fig. 2c in~\cite{Marcus2020}). Current-bias measurements in the SQUID geometry show a trivial Josephson effect in both lobes.  
In the odd-occupied regime a bright zero-bias peak develops at the closing of the zeroth lobe (around half flux quantum), which the authors identify as a Kondo peak. In the destructive regime no peak is visible (superconductivity is fully suppressed in the shells). The zero-bias feature reappears in the first lobe, but less bright and a bit broadened. In the current-bias measurement a $\pi$ phase is absent in the odd-occupied regime in the first lobe (Fig. 4 in~\cite{Marcus2020}), while it is present in the zeroth lobe (supercurrent reversal). As it was theoretically predicted, a full-shell nanowire could potentially acquire topological properties in the first lobe~\cite{Glazman_Marcus2020}, which could explain the absence of the $\pi$ phase and zero-bias peak by hybridization of a dot electron with MBSs~\cite{Zheng2015,Oppen2017,Fu2018,Black-Schaffer2019,Flensberg2020}. However, current-bias measurements in SQUID geometry show the absence of the $\pi$ phase in the center of the odd sector already at the closing of the zeroth lobe (data in supplemental material of~\cite{Marcus2020}, Fig. S8). An enhancement of the supercurrent in the odd state is clearly visible in comparison to the even state, which is a typical feature of the Kondo effect in S-QD-S junctions~\cite{Belzig2004,Egger2004,Meden2008}. And all these features are qualitatively similar to the ones observed in the first lobe (nicely captured in Figs. 4e-f in~\cite{Marcus2020}): the CPR of the SQUID is given by a sinusoid, but in the odd state the critical current is larger (higher average value), no phase shift is observed. That suggests that the observed effects are of the same origin. As we have shown in the previous section the superconducting phase winding introduces a new important energy scale - $\Delta_{eff}$, which plays the role of the effective gap for the QD and which is reduced in comparison to $|\Delta|$ due to destructive interference. In case of nonzero phase winding $n>0$ off-diagonal elements of $\Sigma^{S}$ [see Eq.~(\ref{eq:ssigma})] are smaller by a factor of $\Delta_{eff}/\Delta$, which suggests that in such a system the relevant parameter for the quantum phase transition between the unscreened doublet to the many-body Kondo singlet ground states is $\Delta_{eff}/T_K$. A more formal way to see that is to perform a renormalization group (RG) analyses: the RG flow starts at large energy cutoff, off-diagonal terms of $\Sigma_{\sigma}^{U}$ get renormalized due to off-diagonal elements of $\Sigma^{S}$, see Appendix~\ref{appendix:FRG}. As a result, we were able to deduce that in the first lobe for odd occupation of the QD $T_K>\Delta_{eff}$ and the ground state is a Kondo singlet, which explains the $0$ phase behaviour and the supercurrent enhancement as well as zero-bias peak in the differential conductance. Another question arising is whether the Zeeman effect can play a significant role, because the Zeeman field can split the Kondo peak if the g-factor is large enough~\cite{Meir1994,Kastner1998,Mora2018}. However, it was shown that due to spin-orbit interaction the effective g-factor on a long QD can be renormalized (towards small values)~\cite{Golovach2008,Dmytruk2018,Bosco2021}. That significantly complicates theoretical comparison of Zeeman energy $V_Z=\mu_BgB$ and Kondo temperature $T_K$. In Appendix~\ref{appendix:peak} we provide some simple analyses of the experimental data. 

Another distinctive feature observed in experiment~\cite{Marcus2020} is a change of dissipation between zeroth and first lobes: one can see a strong hysteresis in supercurrent though the SQUID in the zeroth lobe, which indicates underdamped junction regime corresponding to low dissipation. On contrary, in the first lobe no hysteresis is seen; this effect is independent of the QD occupation, therefore, it is not caused by the Kondo effect itself. We suggest that the higher damping can be attributed to lower effective gap (and described in terms of subgap states induced by the vortex~\cite{Vaitiekenas2022}). The different junction dissipation regime in two lobes also implies a different ratio of critical and switching current. In the underdamped regime the switching current (which is measured in the experiment) can be significantly lower than the actual critical current, as the macroscopic quantum phase tunneling cannot be neglected, while in the first lobe strong dissipation (overdamped regime) should result in the switching current being in good correspondence with the critical current.

The experimental setup~\cite{Marcus2020} appears to be a very convenient and unique device to study competition between superconductivity and the Kondo effect at relatively low magnetic fields without additional gates to control the tunneling amplitude due to destructive Little-Parks effect. The regime of competition is specifically interesting to study experimentally as no analytical approach exists to provide quantitative description of the system in this regime. The setup allowed the scientists to measure the CPR and differential conductance in the middle of the odd occupation sector all the way from the doublet ground state to the Kondo singlet smoothly varying the superconducting gap by changing the magnetic flux from zero to a half flux quantum, which does not even require going into the first lobe. The well resolved CPR close to the phase transition (at $40\,mT$ and $45\,mT$ for the first device; Fig. S8 in~\cite{Marcus2020},) has a drastic difference, which is an outstanding feature of the change in the character of the ground state and is in perfect qualitative agreement with theoretical predictions (numerics). We suggest that a measurement of the CPR at different values of flux with smaller steps around the transition could be sufficient to establish the transitions between $0$, $0^\prime$, $\pi^\prime$ and $\pi$ phases of such a QD-based junction~\cite{Arovas1999,Yeyati2003} in the middle of the odd parity sector (before such transitions have been observed only as a function of gate voltage~\cite{Deblock2015,Deblock2018}). For this we recommend to fabricate an asymmetric SQUID with an ancilla junction being independent of the flux through the shell (i.e. a separate SIS junction) and having slightly larger critical current so that the CPR of S-QD-S junction is directly observed (large difference in critical currents between the junctions forming the SQUID decreases the contrast of the picture). Further study of the first lobe can provide better understanding of the Kondo cloud formation due to destructive interference. Moreover, it could be interesting to study the effect in shells of larger radius (or thinner shell, so that the superconducting coherence length is shorter than the shell's radius~\cite{Marcus2020_1}), when Little-Parks effect does not suppress superconducting gap to zero. In that regime suppression of the gap at half flux quantum may not be enough to enhance the Kondo effect, however, the phase winding at higher magnetic fields can still reduce the effective gap to the values below the Kondo temperature, which would result in a zero-bias peak only in the first (or higher) lobe.

\section{Conclusions and outlook}
We provided a coherent interpretation for the results observed in an experiment~\cite{Marcus2020} on a transport through a QD between two full-shell nanowires. Due to accurate and sufficient data presented, we were able to establish the effect of superconducting order parameter phase winding on a ground state of the QD and attribute it to the Kondo effect. We showed that the qunatum phase transition between a doublet and a many-body Kondo singlet ground state is controlled by a parameter $\Delta_{eff}/T_K\ll|\Delta|/T_K$ in case of a superconducting phase winding. We discussed the consequences of this transitions and suggested experiments to test the existing theoretical results on the regime of competition of the Kondo effect and superconductivity. Finally, theoretical analyses of the results suggest that the conductance enhancement due to the Kondo effect in a vortex may as well explain zero-bias anomalies observed in different systems, such as the vortexes localized at magnetic impurities of some superconductors~\cite{Gao2018,Tamegai2019}, which requires further studying.

{\it Acknowledgements.} We thank Wolfgang Belzig, Mikhail Pletyukhov, Charles Marcus and Saulius Vaitiekėnas for fruitful discussions. This project received funding from the European Union's Horizon 2020 research and innovation program (ERC Starting Grant, grant agreement No 757725).

\bibliographystyle{apsrev4-2}
\bibliography{QD}

\appendix
\section{Effective gap}\label{app:effdelta}

The effective Hamiltonian~(\ref{eq:hamiltonian}) was derived in~\cite{Prada2022} within the SIA model and consists of three parts: bare QD energy, Coulomb interactions, and proximity effect from the shell. For a usual S-QD-S junction (no phase winding) the latter is given by~\cite{Yeyati2003,Meden2008}
\begin{equation}\label{eq:proximity}
\Sigma^{S}=-\sum_{s}\frac{\pi\rho_{S}t^2}{\sqrt{\Delta^{2}-\omega^{2}}}\left(\begin{array}{cc}
\omega  & \left|\Delta\right|e^{i\phi_{s}}\\
\left|\Delta\right|e^{-i\phi_{s}} & \omega 
\end{array}\right),
\end{equation}
where the index $s$ stands for left/right superconductor (we assume $\left|\Delta\right|$ and $t$ to be the same for the left/right shells). As discussed in~\cite{Prada2022}, the approach can be extended to the hollow-cylinder model, which allows us to include a magnetic flux in the model. Here we briefly sketch the main steps of that approach, the difference with~\cite{Prada2022} being that here we consider superconducting shells both to the right and to the left of the QD, therefore, we need to take a relative phase into account as well. One needs to sum over all the possible positions on the left/right shell, which is done by introducing angle-dependent tunneling amplitudes $t(\theta)/(2\pi)$ as well as order parameters $\Delta_s=|\Delta|e^{i\phi_s-in\theta}$, where $n$ is the number of the phase windings $n=\lfloor\Phi/\Phi_0\rceil$ due to the flux through the shell, see Fig.~\ref{fig:SQDS}. As was reported in~\cite{Prada2022} for the axially symmetric case $t(\theta)=const$, the off-diagonal elements for any $n\neq0$ are zero, which can be seen as destructive interference. Realistically, the wave function of an electron on the QD cannot be treated as axially symmetric (with respect to the shell axis) due to gating from one side and due to inhomogeneities. However, we can still assume the tunneling to have a relatively weak angle dependence (as discussed in~\cite{Prada2022} the tunneling is determined by the wave-function's tails underneath the shell, which are not very sensitive to the gating): $t(\theta)=t_0+\delta t(\theta)$. It is convenient to decompose this tunneling amplitude into harmonics $t(\theta)=\sum_mt_me^{-im\theta}$, where $t_m=\int_{-\pi}^\pi t(\theta)e^{im\theta}\frac{d\theta}{2\pi}$. Then, one can get the proximity term in the form of Eq.~(\ref{eq:proximity}) by integrating the shell contribution over the angle $\theta$. The simplest estimation comes from expanding $t^2(\theta)=t_0^2+2t_0\delta t(\theta)+...$.  The off-diagonal term takes the form
\begin{equation}\label{eq:winding}
\Sigma^{S}_{01}\approx-\sum_{s}\frac{2\pi\rho_{S}t_{0}\left|\Delta\right|e^{i\phi_s}}{\sqrt{\Delta^{2}-\omega^2}}\intop_{0}^{2\pi}\frac{t(\theta)e^{-in\theta}d\theta}{2\pi}.
\end{equation}
Let us derive the term in a more accurate fashion. The thin superconducting shell Hamiltonian can be written in cylindrical coordinates as~\cite{Glazman_Marcus2020,Prada2020}
\begin{multline}
H_s=\left[\frac{k_s^2+k_r^2+\left(k_\theta+eA\tau_z\right)^2}{2m_s}-\mu_s\right]\tau_z+\\
+|\Delta|\left(\cos\left[-in\theta+i\phi_s\right]\tau_x+\sin\left[-in\theta+i\phi_s\right]\tau_y\right),
\end{multline}
where $k_s$, $k_r$ and $k_\theta$ are the longitudinal, radial, and tangential components of the momentum operator, $m_s$ is the effective electron mass, $\mu_s$ is the chemical potential in the shell; $A=\frac{1}{2eR}\frac{\Phi}{\Phi_0}$ is the vector potential, $\tau_i$ are Pauli matrices acting in Nambu-Gorkov space. One can introduce a generalized angular momentum $J_z=-i\partial_\theta+\frac{1}{2}n\tau_z+\frac{1}{2}\sigma_z$, which is conserved (commutes with Hamiltonian)~\cite{Glazman_Marcus2020,Vaitiekenas2022}, then the eigenvalues $m_J$ are good quantum numbers and can take half-integer values:
\begin{equation}
\left[m_J-\frac{1}{2}-\frac{1}{2}n\right]\in \mathbb{Z}.
\end{equation}
The resulting angular momentum number $m$ is integer and fixed by $m_J$ and $n$ in each spin and Nambu sector (note that it is exact only in the absence of radial spin-orbit interaction~\cite{Vaitiekenas2022}). Due to the  term $\frac{1}{2}n\tau_z$ in the definition of the generalized angular momentum, for a fixed spin the Hamiltonian is not block-diagonal in the $(m,-m)$ Nambu sectors, but in the
$(m,-m-n)$ sectors~\cite{Prada2022}. 
 The retarded Green function for the shell (decoupled from the QD) between two positions $\theta$ and $\theta^\prime$ is then given by~\cite{Prada2022}
\begin{widetext}
\begin{equation}
g^s(\omega,\theta,\theta^\prime)=\sum_{m}\frac{e^{-im(\theta-\theta^\prime)}}{D_{m,n}}\left(\begin{array}{cc}
\omega+\epsilon_{k_s}+\frac{\left[m+n-\Phi/(2\Phi_{0})\right]^{2}}{2m_sR^{2}} & \Delta e^{-i\phi_s+in\theta^\prime}\\
\Delta e^{i\phi_s-in\theta} & \left[\omega-\epsilon_{k_s}-\frac{\left[m+\Phi/(2\Phi_{0})\right]^{2}}{2m_sR^{2}}\right]e^{-in(\theta-\theta^\prime)}
\end{array}\right),
\end{equation}
where 

\begin{equation}
D_{m,n}=\omega^2-\Delta^2-\left(\epsilon_{k_s}-L_{m}\right)^2
-\left(\omega-\epsilon_{k_s}+L_{m}\right)\left(\Phi_n-2\sqrt{\Phi_n L_{m}}\right),
\end{equation}
with
\begin{equation}
L_{m}=\frac{\left(m+\frac{\Phi}{2\Phi_0}\right)^2}{2m_sR^2}\quad\mathrm{and}\quad \Phi_n=\frac{\left(n-\frac{\Phi}{\Phi_0}\right)^2}{2m_sR^2}.
\end{equation}
\end{widetext}
Then the proximity effect of the shells on the QD can be described by the self-energy~\cite{Prada2022}
\begin{equation}
\Sigma^S(\omega)=\sum_{s}\int dk_s\int \frac{d\theta}{2\pi} \frac{d\theta^\prime}{2\pi} t(\theta)g^s(\omega,\theta,\theta^\prime)t(\theta^\prime).
\end{equation}
In the middle of the first lobe $\Phi/\Phi_0=n=1$ the result can be written in the form of Eq.~(\ref{eq:proximity}) but with $t$ replaced by $t_0=\int_{-\pi}^\pi t(\theta)\frac{d\theta}{2\pi}$, which is just the tunneling amplitude averaged over $\theta$, and
 $|\Delta|$ in the nnumerator of the off-diagonal terms replaced by 
\begin{equation}
\Delta_{eff}=\left|\sum_{m}\frac{t_{-m}t_{m+n}}{t_0^2}\Delta\right|.
\end{equation}
 In case of weak modulation compared to the angle-independent tunneling ($t_0\gg|\delta t(\theta)|$) one gets for $n\neq 0$
\begin{equation}\label{eq:deltaeff}
\Delta_{eff}\approx2\left|\frac{\Delta\intop_{-\pi}^{\pi}t\left(\theta\right)e^{i\theta n}\frac{d\theta}{2\pi}}{\intop_{-\pi}^{\pi}t\left(\theta\right)\frac{d\theta}{2\pi}}\right|\ll\left|\Delta\right|,
\end{equation} 
which is the same as Eq.~(\ref{eq:winding}) [if one substitutes $\Delta_{eff}$ for $\Delta$ in the numerator of the off-diagonal terms of Eq.~(\ref{eq:proximity})].

Away from the first lobe center ($\Phi/\Phi_0\neq1$) the off-diagonal elements have additional terms in the denominator:
\begin{equation}
\Sigma^{S}_{01}=-\sum_m\frac{2\pi\rho_{S}t_{-m}t_{m+n}\left|\Delta\right|\cos\frac{\phi}{2}}{\sqrt{\Delta^{2}-\left(\omega+\frac{(n/2+m)(n-\Phi/\Phi_0)}{2m_sR^2}\right)^2}}.
\end{equation}
However, these terms do not change the qualitative picture, therefore, Eq.~(\ref{eq:winding}) gives a reasonable estimation, which makes $\Delta_{eff}$ given by~(\ref{eq:deltaeff}) a relevant energy scale instead of $|\Delta|$.

The spin-orbit interaction was not taken into account in the simplified hollow-cylinder model, as it can play a role only in the semiconducting nanowire itself, therefore, it should not be crucial for the shell modelling and can only affect tunneling process to the QD states (introducing small spin-dependent corrections to the tunneling amplitudes~\cite{Nazarov2009}) and QD parameters. For the latter we can use empirical effective parameters, i.e. in the main text we stated that the effective Zeeman splitting is reduced due to spin-orbit interactions so that the effective $g$ factor is small~\cite{Golovach2008,Dmytruk2018,Bosco2021}. Therefore, we claim that including spin-orbit interactions as well as more realistic (i.e. hexagonal) geometry into consideration should not affect the result qualitatively.

\section{Analysis of Kondo conductance peak}\label{appendix1}\label{appendix:peak}
Here we provide a simple estimate from below on the Kondo temperature based on data provided in~\cite{Marcus2020} and supplemental material therein. From the supplemental material (Fig. S7) we can see at $45\,\mathrm{mT}$ a Kondo enhancement in the odd state (larger supercurrent amplitude in the odd state in comparison to the even state and no $\pi$ phase shift), while at $40\,\mathrm{mT}$ we see a $\pi$ phase behaviour and similar supercurrent amplitudes in odd and even sectors. As a result, we can crudely estimate the Kondo temperature from below as $T_K\gtrsim\Delta$ at $45\,\mathrm{mT}$. NRG calculations predict a transition from a doublet ground state to a many-body Kondo singlet at $T_K\approx 0.3\Delta$~\cite{Shimizu1992,Satori1993,Hewson2007}, however, it was defined at zero phase difference, while the current-phase relation at $45\,\mathrm{mT}$ already shows a critical current enhancement as well as $0$ phase behaviour at all the phases, for nonzero phase the transition occurs at higher values of $T_K/\Delta$~\cite{Novotny2016,Meden2019}, which suggests that the system is already deep in the Kondo regime. The gap is suppressed at $50\,\mathrm{mT}$, we thus can use a simplified formula for a thin shell of radius $R$~\cite{Wang2001,Moler2007}):
\begin{equation}
\Delta(\Phi)\approx\Delta_0\max\left\{1-\frac{\xi^2}{R^2}\left(n-\frac{\Phi}{\Phi_0}\right)^2,0\right\},
\end{equation}
where $\xi$ is the superconducting coherence length, $n=\lfloor\Phi/\Phi_0\rceil$  the number of phase windings in the shell ($n=0$ for the case here, as the system is still in the zeroth lobe). Therefore, for an estimate we take $1-\frac{\xi^2}{R^2}\left(-\frac{50}{120}\right)^2=0$, as the center of the first lobe is at $B=120\,\mathrm{mT}$, which corresponds exactly to one flux quantum. Then at $45\,\mathrm{mT}$
\begin{equation}
\Delta(\Phi)\approx\Delta_0\left[1-\frac{12^2}{5^2}\left(\frac{45}{120}\right)^2\right]=0.19\Delta_0.
\end{equation}
The order parameter in the Al shell may be estimated from below from the differential conductance at zero magnetic field: $\Delta_0>0.1\,\mathrm{meV}$. Then, we can estimate the Kondo temperature from below: $T_K>19\,\mu\mathrm{eV}$; one can see that this estimate from below gives a value twice larger than originally assumed. We stress again that the estimate is crude and well below the real value, a more accurate evaluation of the Kondo temperature could be possible with more data around the transition from the doublet to the Kondo singlet ($40-45\,\mathrm{mT}$) or the Kondo peak analysis in the destructive regime (normal state).  

Data on the second device provided in the supplemental material of Ref.~\cite{Marcus2020} shows similar features suggesting an enhancement of the Kondo effect. Moreover, in Fig. S2 the differential conductance peak at low bias voltage appears to be split in the first lobe, which may be due to Zeeman splitting of the Kondo peak. Fig. S4 shows that the critical current in the ancilla junction is larger in the first lobe than in the QD-based junction (the amplitude of the supercurrent in the SQUID is changing with the QD occupation), which suggests that the tunneling amplitude in the second device could be smaller than in the first one. Therefore, we expect that the Kondo temperature in the second device is lower (or the effective $g$-factor can be larger) and the Kondo peak acquires a visible splitting in the magnetic field ($2V_Z>T_K$ in the first lobe), which cannot be explained by MBSs. Furthermore, all three devices show supercurrent enhancement in the odd state (first lobe) in comparison to even occupation, which is also a typical feature of the Kondo effect. 
In conclusion, the additional data on the second and third devices in the supplemental material of Ref.~\cite{Marcus2020} supports our idea of an enhancement of the Kondo effect in the first lobe.

\section{QD interactions}\label{appendix:FRG}
The self-energy due to interactions can be fully covered only by elaborate numerical methods, such as NRG~\cite{Satori1993,Ohashi2000,Hewson2007,Yeyati2012}. However, some approaches can give at least qualitative insights in the system behaviour at different values of $T_K/\Delta$. For example, in~\cite{Meden2008} the author proposed an FRG method~\cite{Meden2006}, which gives within the lowest-order static approximation flow equations from high energy cutoff $\Lambda$ in the Matsubara non-interacting Green-function $G^{0,\Lambda}(i\omega)=\theta(\omega-\Lambda)G^0(i\omega)$ in the form
\begin{widetext}
\begin{equation}
\partial_{\Lambda}\Sigma_{01}^{U}(\Lambda)=-\frac{U(\Lambda)}{\pi}\frac{\Sigma_{01}^{U}(\Lambda)-\Delta_{eff}\frac{2\langle\Gamma_{s}\rangle\cos\frac{\phi}{2}}{\sqrt{\Lambda^{2}+\left|\Delta\right|^{2}}}}{D^\Lambda(i\Lambda)},
\end{equation}

\begin{equation}
\partial_{\Lambda}U(\Lambda)=-\frac{2}{\pi}\left[\frac{U(\Lambda)}{D^\Lambda(i\Lambda)}\right]^2\left|\Delta_{eff}\frac{2\langle\Gamma_{s}\rangle\cos\frac{\phi}{2}}{\sqrt{\Lambda^{2}+\left|\Delta\right|^{2}}}-\Sigma_{01}^{U}(\Lambda)\right|,
\end{equation}
\begin{equation}
D^\Lambda(i\omega)=\omega^{2}\left(1+\frac{2\langle\Gamma_{s}\rangle}{\sqrt{\omega^{2}+\left|\Delta\right|^{2}}}\right)+\left|\Delta_{eff}\frac{2\langle\Gamma_{s}\rangle\cos\frac{\phi}{2}}{\sqrt{\omega^{2}+\left|\Delta\right|^{2}}}-\Sigma_{01}^{U}(\Lambda)\right|^{2}.
\end{equation}
\end{widetext}
The equations are for the middle of the odd sector (which implies trivial diagonal elements) and symmetric (left/right) tunneling, $\Sigma_{01}^{U}(\infty)=0$, $U(\infty)=U$. In case of no phase winding, which was studied in~\cite{Meden2008}, $\Delta_{eff}=\left|\Delta\right|$ and $\langle\Gamma_s\rangle=\Gamma/2$, while in our case (the first lobe), the effective gap is reduced, which implies that the flow is slow and the off-diagonal term $\Sigma_{01}^{U}(\Lambda)$ does not reach the critical value for the quantum phase transition into the $\pi$ phase, which can be seen from the expression for the supercurrent~\cite{Meden2008}:
\begin{widetext}
\begin{equation}\label{eq:current}
\langle J\rangle=\frac{1}{2\pi}\int\left[\frac{\langle\Gamma_S\rangle^2\Delta^2_{eff}\sin\phi}{D^{\Lambda=0}(i\omega)(\omega^2+\left|\Delta\right|^{2})}-\frac{2\langle\Gamma_S\rangle\Delta_{eff}\Sigma_{01}^{U}(0)\sin(\phi/2)}{D^{\Lambda=0}(i\omega)\sqrt{\omega^2+\left|\Delta\right|^2}}\right]\frac{d\omega}{2\pi}.
\end{equation}
\end{widetext}
If the first term dominates after the renormalization of $\Sigma_{01}^{U}$, which is the case of low $\Delta_{eff}$, the junction is in the $0$ phase, otherwise in the $\pi$ phase. One should note that Eq.~(\ref{eq:current}) is exact, the $0$ or $\pi$ phase behaviour of the QD-based junction is determined by the ratio of two competing terms, however, it is $\Sigma_{01}^{U}(0)$ in the second term which cannot be calculated exactly in the limit of strong interactions.

\end{document}